# High Speed Friction Microscopy and Nanoscale Friction Coefficient Mapping

James L. Bosse, Sungjun Lee, Bryan D. Huey*, Andreas Sø Andersen, Duncan S. Sutherland

J. L. Bosse, Dr. S. Lee, Prof. B. D. Huey
Department of Materials Science & Engineering
97 North Eagleville Road, Unit 3136
Storrs, CT 06269-3136
E-mail: bhuey@ims.uconn.edu

A. S. Andersen, Prof. D. S. Sutherland
Interdisciplinary Nanoscience Center (iNANO)
Gustav Wieds Vej 14
Aarhus University
DK-8000 Aarhus C

*As mechanical devices in the nano/micro length scale are increasingly employed, it is crucial to understand nanoscale friction and wear especially at technically relevant sliding velocities. Accordingly, a novel technique has been developed for Friction Coefficient Mapping (FCM), leveraging recent advances in high speed AFM. The technique efficiently acquires friction versus force curves based on a sequence of images at a single location, each with incrementally lower loads. As a result, true maps of the coefficient of friction can be uniquely calculated for heterogeneous surfaces. These parameters are determined at a scan velocity as fast as 2 mm/s for microfabricated $SiO_2$ mesas and Au coated pits, yielding results that are identical to traditional speed measurements despite being ~1000 times faster. To demonstrate the upper limit of sliding velocity for the custom setup, the friction properties of mica are reported from 200 µm/sec up to 2 cm/sec. While FCM is applicable to any AFM and scanning speed, quantitative nanotribology investigations of heterogeneous sliding or rolling components are therefore*

*uniquely possible, even at realistic velocities for devices such as MEMS, biological implants, or data storage systems.*

## 1. Introduction

Literally for centuries [1], studies have been conducted into friction, the related phenomena of wear, adhesion, and lubrication, and ultimately materials and component design to optimize sliding or rolling performance. Such tribological investigations are especially relevant to micro- and nano-electromechanical systems (MEMS/NEMS), which despite their widespread application in accelerometers, DLP projectors, ink-jet or fuel-injector heads, etc., can be hampered due to the relatively high adhesion forces at such small length scales [2-8]. The study of nanotribology aims to characterize, understand, and control these effects, and is principally conducted with variations of atomic force microscopy (AFM) [9, 10]. However, MEMS/NEMS operate with velocities of tens of millimeters per second or more, significantly faster than the speed of most AFM systems [10]. Moreover, as device complexity continues to increase, heterogeneities in the local friction response are also increasingly relevant, but are difficult to quantify using present methods. Accordingly, this work is concerned with the development and application of a quantitative friction mapping method [11, 12], operating at technically relevant sliding velocities, and suitable for real, heterogeneous surfaces.

AFM-based nanotribology is primarily accomplished using lateral force measurements, acquired by monitoring lever torque during contact-mode scanning perpendicular to the lever axis. So called Lateral Force Microscopy (LFM) [13, 14] images are then based on a friction signal calculated at any given position from the difference in torsional contrast when scanning in opposite directions. This is performed with a fixed normal load, and typically at a low speed of ~1-10 μm/sec, ultimately yielding an LFM image that maps local friction behavior, although the quantitative nature of such images is limited due to the number of fixed variables.

To address this issue [14-19], the same lateral friction force signal is recorded as described above, but also for a range of normal loads approaching loss of contact. The coefficient of friction is then calculated from the slope of the lateral versus normal forces. Other parameters can also be extracted, including the force at zero normal load, the attractive force at zero friction, and any points of discontinuity. However, this approach generally provides a value instead of an entire image, as it is typically based on averaging the friction signal from multiple pixels, scan lines, or image frames, then incrementally changing the force, acquiring LFM data again, etc. It is thus primarily applicable to relatively homogeneous specimens where changes in location do not appreciably influence the adhesion characteristics.

To address the need for quantitative friction coefficient mapping, the current work uniquely leverages a custom high speed SPM system [20, 21] to essentially combine LFM imaging and force-dependent friction measurements. Based on a sequence of high speed images, each acquired in the same area but with incrementally lower applied loads, this efficiently provides a 3-d dataset of friction versus area as sketched in Figure 1. A similar but standard-speed approach apparently developed in parallel was recently reported, though it notably does not measure the actual lateral friction and hence cannot extract friction coefficients [11]. Furthermore, it does not implement high speed imaging, and hence the data density is relatively sparse and does not allow velocity dependent studies over several orders of magnitude as reported here.

The array of friction-force curves which FCM provides, 65,536 of them for standard 256x256 pixel LFM images, thus provides a high data density of friction information for the imaged area. Nanoscale maps of the coefficient of friction, friction at zero load, and/or load at zero friction can therefore be uniquely and efficiently generated, most importantly for surfaces with nanoscale heterogeneities in phases, topography, defects, etc. Tip speeds up to 2 cm/s are specifically considered with results that agree with previously reported models and experiments, demonstrating nanotribology investigations over several orders of magnitude of sliding velocity.

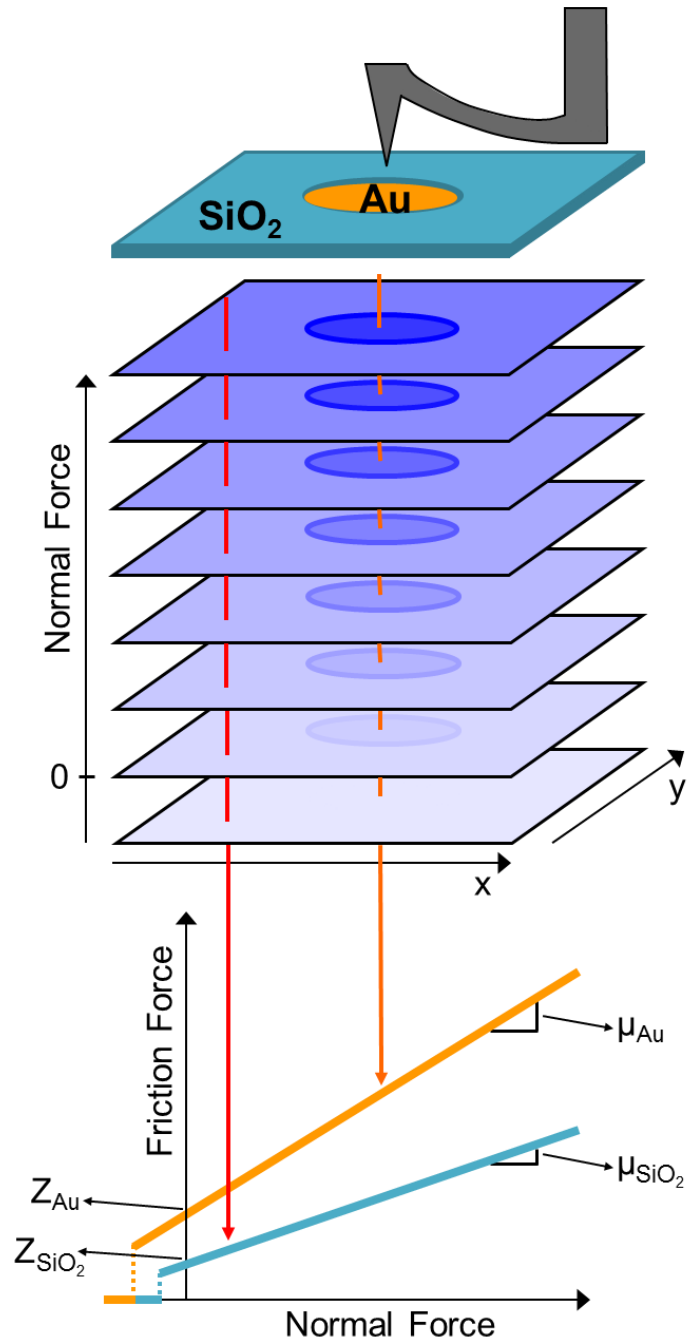

Figure 1: Sketch of the high speed Friction Coefficient Mapping approach for a heterogeneous $SiO_2$/Au specimen. Lateral force microscopy images are acquired with incrementally lower applied loads until loss of contact occurs. Friction force curves are then extracted for each pixel to quantify local friction properties.

## 2. Standard Speed Friction Measurements

Figure 2(a) typifies a common application of AFM based friction studies. As revealed by an SEM image, Figure 2(b), the specimen is a nanopatterned surface fabricated by sparse colloidal lithography. This surface has pits in a 14nm thick electron beam evaporated $SiO_2$ layer, revealing 300nm diameter circular patches of an underlying granular thin film of Au deposited on a silicon wafer (substrate).

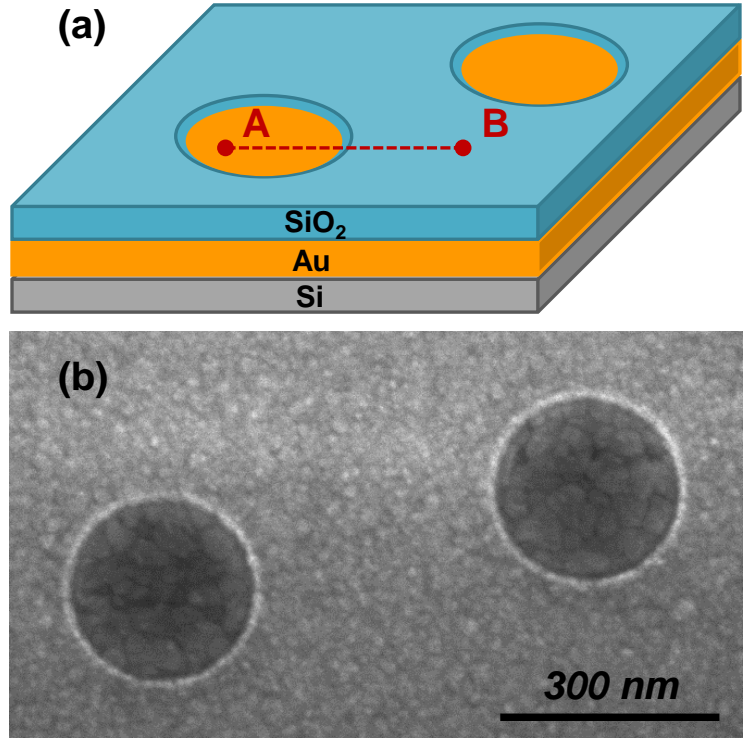

**Figure 2. Sketch (a) and SEM image (b) of a model nanostructured specimen with Au bottomed pits in a SiO2 film.**

The AFM based topography for such a specimen, Figure 3(a), mimics the SEM contrast. Simultaneously acquired qualitative friction images, i.e. normal speed (4 µm/sec) LFM images, expectedly display contrast due to the distinct materials (Au and $SiO_2$) as well as edge effects due to the topographic step, Figure 3(b). Nanostructuring of the friction response for the $SiO_2$ layer is also apparent, with feature dimensions as small as 10 nm clearly resolved related to the nanoscale grains of the polycrystalline $SiO_2$ film. The grain structure of the sputter deposited Au film is visible at the bottom of the pits as well. Of course, intermittent or non-contact AFM based phase imaging can provide similar qualitative images of friction, as it too can relate to local adhesion. Truly measuring the friction coefficient throughout the imaged area is challenging, however, especially for nanostructured surfaces where the roughness and variability hinders local friction quantification.

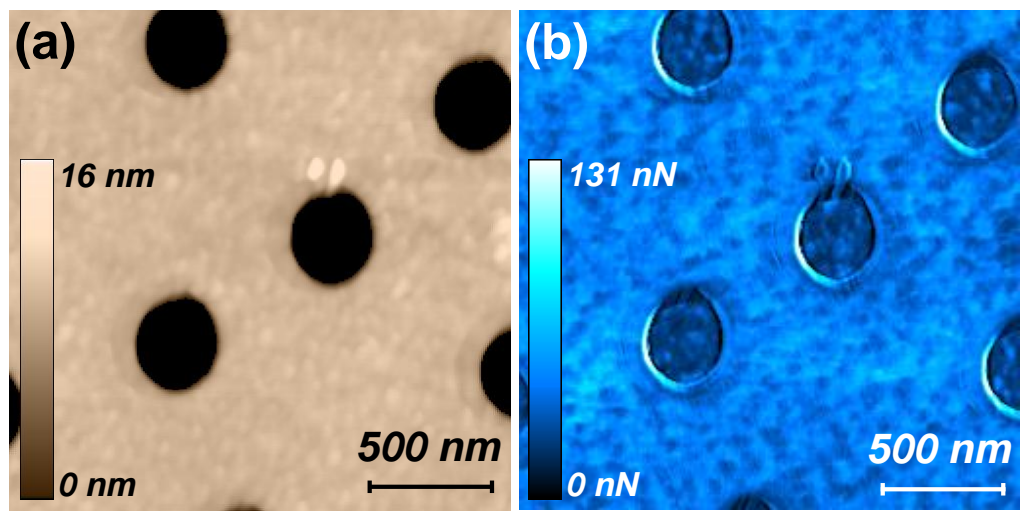

**Figure 3. Standard speed AFM (a) and LFM (b) images (2 µm x 2 µm, 4 µm/s sliding velocity) of Au coated pits surrounded by $SiO_2$.**

The traditional approach to quantifying lateral force data is to continuously re-scan a single line traversing multiple phases, e.g. between points A (Au) and B (SiO$_2$) as sketched in Figure 2(a), but crucially for a range of normal loads. Friction-force curves for each region are then typically generated by plotting the response from the distinct regions [17, 18], e.g. the left (Au) and right (SiO$_2$) halves of the repeated friction lines at the corresponding normal forces. Accordingly, Figure 4 presents the lateral friction with normal loading from 530 nN down to 2.75 nN, based on 96 force steps (lines of data) of -5.55 nN each applied along a single 500 nm line. Overlain error bars indicate a negligible standard deviation in the measured lateral force for each normal load. This implies a relatively uniform friction response in the distinct sample regions, at least within the 200-250 nm linear regions of each phase that were sampled. Standard deviations for the normal force error are too insignificant to see (<1%).

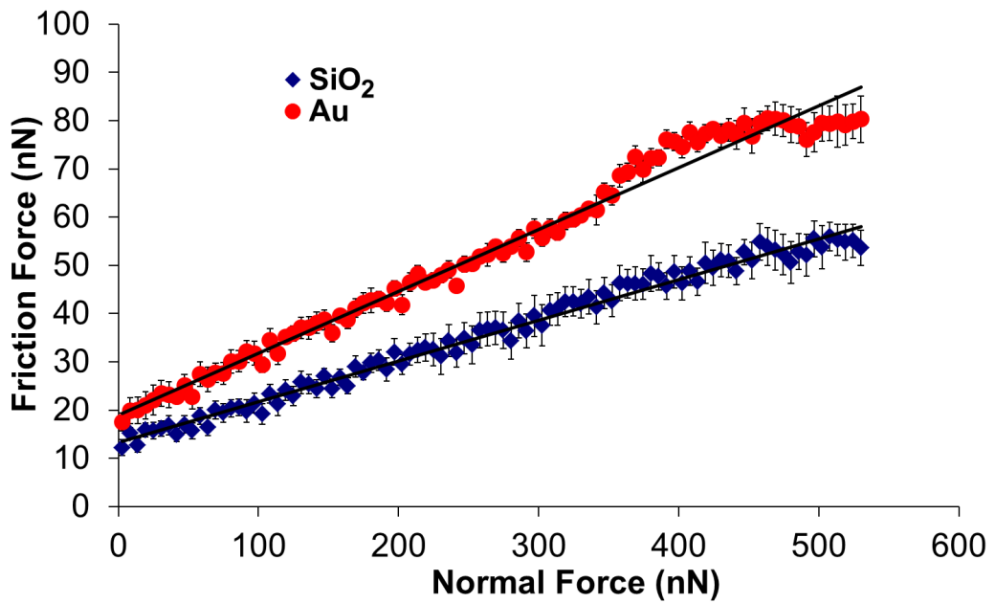

**Figure 4.** Friction force curves for SiO$_2$ and Au acquired at standard speeds (10 µm/s) for comparison with high speed results (standard deviation error bars are shown). The coefficient of friction for Au and SiO$_2$ over **the entire loading range is 0.13 ± 0.01 and 0.08 ± 0.02, respectively, while the friction at zero applied force is 19 ± 4 nN and 13 ± 5 nN.**

The friction coefficient (slope), and friction at zero load (offset), are clearly distinguishable for the two specimen regions. Based on least squares fitting of the entire dataset (overlaid) the Au phase exhibits a higher coefficient of friction than the $SiO_2$ phase (0.13 versus 0.08), with a ratio of the mean values of 1.52:1. Au also exhibits a higher friction at zero applied force with a mean ratio of 1.42:1.

The nonlinearity between 350 and 500 nN for Au is unexpected, and deserves future study particularly to assess the possibility that this indicates some sample wear at such high loads. Even so, however, in this instance it happens to be the case that fitting the purely linear region of the Au curve (from 0 to ~350 nN of normal force only) leads to identical coefficients of friction (within significant digits) as for the entire loading range (0 to ~550 nN normal force). The linear fit $R^2$ values are also equivalent, with 0.98 and 0.99 considering the entire loading range for Au and $SiO_2$ respectively, as compared to 0.99 and 0.98 for the loading range from 0 to 350 nN.

The obvious disadvantage to this line-by-line approach is that subtle variations in friction along the measured line may correlate to specific regions, phases, structures, etc., instead of simply expanding the error bars. Naturally the friction data could be separated into more than 2 subsets, which is conceptually identical and certainly feasible. However, generally this would be impractical as it requires unique computational solutions for any given region of a sample. Moreover, it presumes the ability to distinguish the unique regions, a particular challenge along just 1 dimension. Finally, the method is susceptible to position drift as well, with the tip practically scanning a slowly shifting line, contributing to possible load-dependent error for inhomogeneous surfaces or in the event of wear. The ability to image the friction coefficient is therefore paramount.

## 3. Friction Coefficient Mapping

The results presented above implemented a relatively standard sliding velocity of 10 µm/sec, based on a line rate of 10 Hz. However, measurements at much higher velocities are feasible with high speed SPM, employing line scanning rates on the order of hundreds to thousands of Hz that correspond to sliding velocities approaching cm/s instead of µm/s. This speed enhancement makes it experimentally practical to rapidly acquire multiple images of the friction signal at distinct normal loads as explained for Figure 1, instead of simply detecting friction for a single scan line (or part of one) as in Figure 4. Accordingly, Figure 5 displays a montage of 10 LFM images extracted from a complete sequence of 28 consecutive scans, each with decremented normal loads from 765 nN down to -3 nN as indicated. All are from a single, 1 µm x 1 µm region with a circular pit present near the image center, for the same specimen as considered in Figures 3-4. The crucial distinction is that here, a 1000 Hz line rate was employed throughout. This is 500 times faster than with Figure 3, requiring only 7 seconds for the entire 28-image experiment.

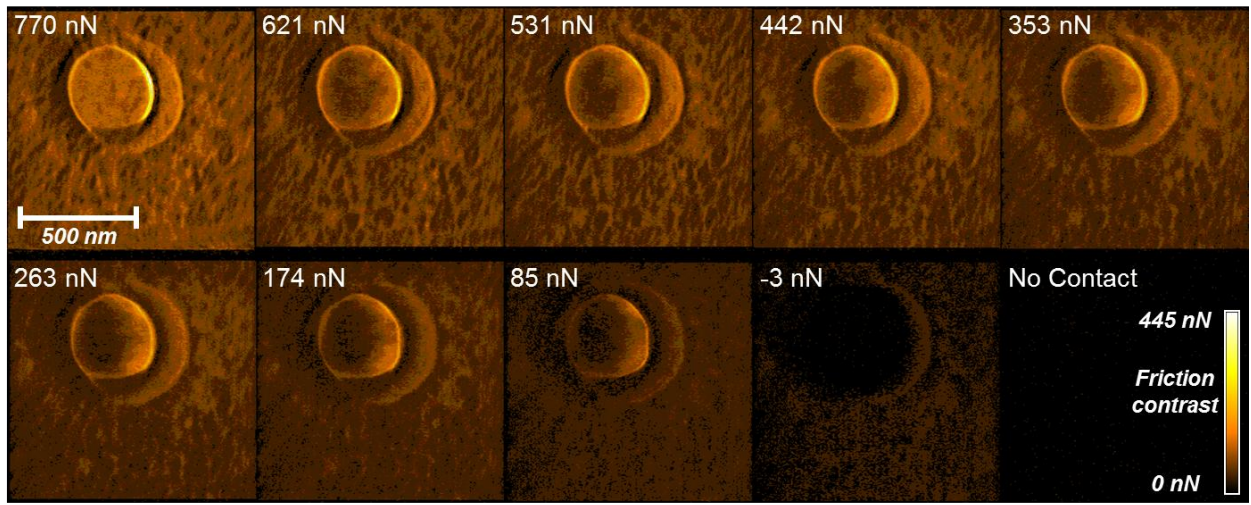

**Figure 5. Montage of high speed friction images at distinct normal loads as labeled, representing a subset of 28 total images for the same 1 μm x 1 μm area, all imaged with a tip velocity of 2000 μm/sec based on a line rate of 1000 Hz.**

Extracting the friction versus normal force from each point in the 250x250 pixel images of Figure 5 therefore can provide up to 62,500 friction-force curves, each similar to those presented in Figure 4. After standard drift correction for the 28 sequential images (forces) as described in the experimental section, the slope of each curve (i.e. for each pixel) can easily be calculated. This is presented in Figure 6(a), a map of the coefficient of friction, with spatial resolution of just 4 nm x 4 nm. Random scatter in each friction versus normal force curve is quantified in Figures 6(b and c), which respectively present the 95% confidence error and the coefficients of determination ($R^2$) for the friction coefficients. Due to the observed nonlinearity in the friction versus normal force curve for Au above ~350 nN during traditional friction measurements (i.e. Figure 4), the FCM-determined coefficient of friction, 95% confidence error and $R^2$ values, are conservatively assessed based only on the linear friction regime (<350 nN).

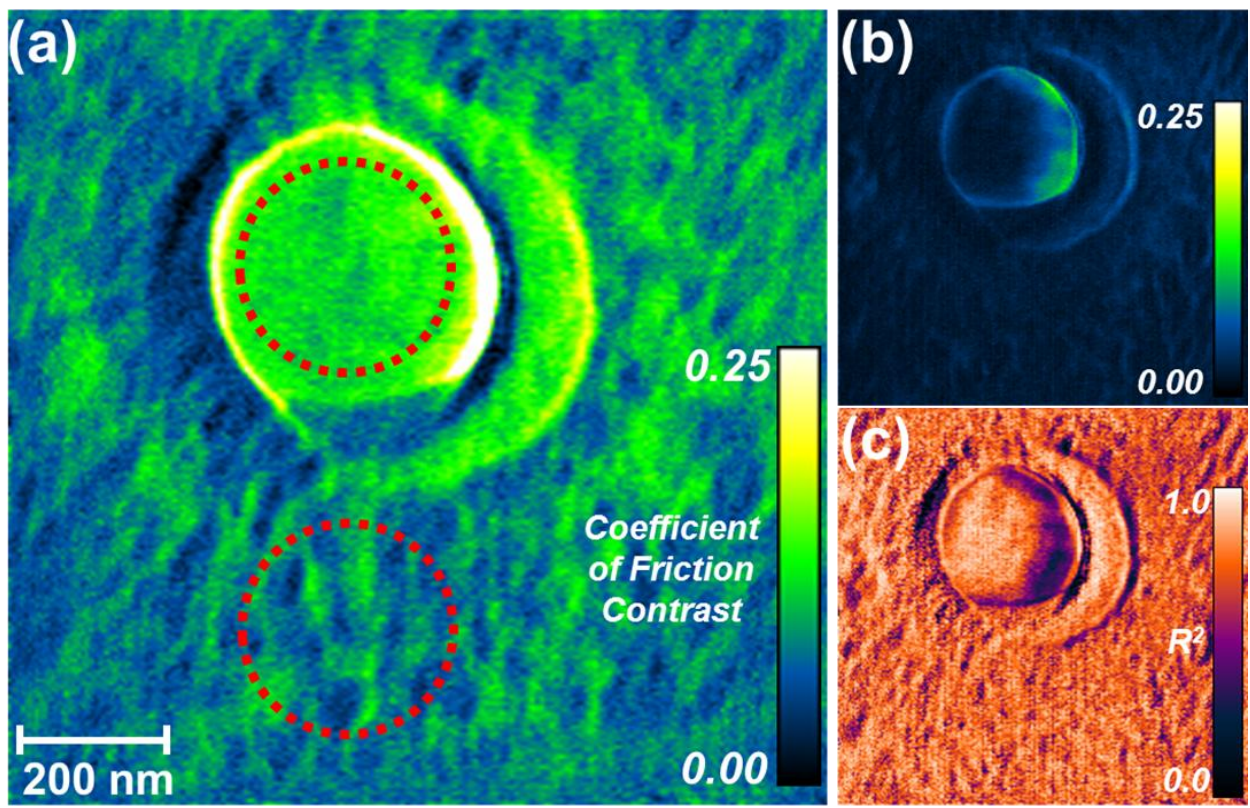

**Figure 6. 968 nm x 956 nm map of the coefficient of friction resolved down to 4 nm x 4 nm, based on 57,838 friction force curves up to normal loads of 350 nN from the dataset of Figure 5, all acquired by high speed SPM in just 7 seconds with a 2 mm/sec tip velocity (a). The corresponding 95% confidence interval (b) and coefficient of determination, $R^2$, (c) of the coefficient of friction map are also shown.**

As with the standard LFM image of Figure 3, higher friction is apparent in Figure 6 for the Au-coated pit-region compared to the surrounding $SiO_2$. Comparing equal areas for these two phases, as sketched in Figure 6, histograms of the results indicate a mean coefficient of friction for the $SiO_2$ region of 0.12 ± 0.01 (standard deviation), while for the Au pits it is 0.08 ± 0.02. But, since the friction coefficient is now mapped, it is also uniquely revealed that friction is more uniform in the pits, as compared to the surrounding $SiO_2$ where nanostructuring is clearly visible (similar to the LFM image of Figure 3). Using traditional nanotribology methods, such heterogeneities due to varying friction and/or adhesion either would have been averaged out (as with Figure 4), caused increased apparent error (standard deviation bars in Figure 4), or more perniciously skewed the results higher or lower than the statistical mode (if the position dependent response were not as symmetric as occurs here, especially in the $SiO_2$).

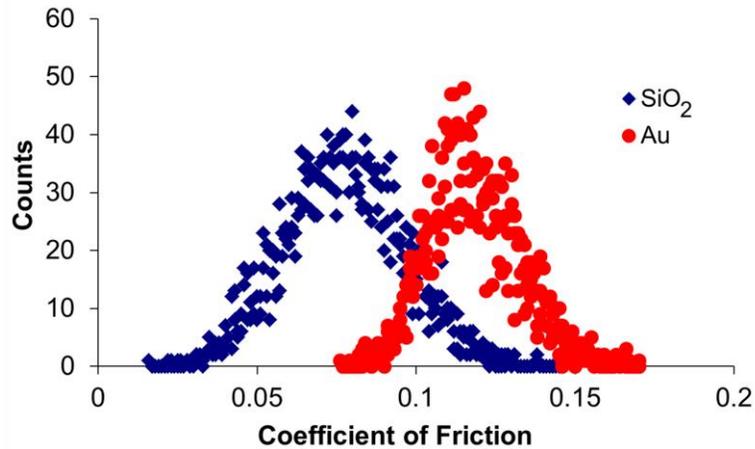

**Figure 7. Histogram of local friction coefficients from equal areas (0.047 μm$^2$) in the two distinct phases of Figure 6.**

It is insightful to compare these high speed results with standard speed friction measurements. The ratio of mean friction coefficients for the Au pit vs. the surrounding $SiO_2$ for the high sliding velocity of ~2 mm/s (from Figure 7) is 1.52:1. At a more common 10 μm/s (200 times slower, from Figure 4) the mean ratio was identical within significant digits. Results from the high speed FCM approach are therefore consistent with traditional speed friction measurements, both visually and quantitatively, with the benefit that they efficiently and spatially resolve the friction.

## 4. Velocity Dependence of Friction

Extending this concept further, high speed SPM can be employed for nanoscale friction investigations at sliding velocities approaching those comparable with actual sliding or rolling applications. Sliding speeds as low as 4 nm/s [14, 22-24] and as high as 200 mm/s have been reported elsewhere [9, 10], while scan lengths from 2 nm [14] to 1 mm [10] have been considered. Studies at these extremes are typically for only a fixed (or just a few arbitrary) load(s), though, instead of the broad range of consecutive loads that are necessary to accurately calculate the friction coefficient and other friction parameters. The highest velocity investigations [9, 10] reported are also for a single scan line (i.e. non-imaging), and hence are not as applicable for heterogeneous surfaces.

Here, the high speed capabilities of the custom SPM system are leveraged to study friction for sliding velocities ranging from ~200 μm/s up to ~2 cm/s. For each speed, the friction force was recorded while scanning at a line rate of 1000 Hz just as in Figure 6, except the scan size was decremented for each new frame, ultimately encompassing 22 distinct sliding velocities. This was performed on a freshly cleaved mica specimen, providing a homogeneous surface exhibiting a few atomic terraces. As before, such topographic features caused variations in the normal load, though again normal forces were simultaneously measured and subsequently employed to calculate the correct local friction contrast. Averaging the results from each image for simplicity (since spatially they are nearly featureless), Figure 8 displays this mean friction signal normalized by the normal load as a function of sliding velocity. A standard deviation of less than

±5% within the area measured was found at every velocity, accounting for the variations in normal load.

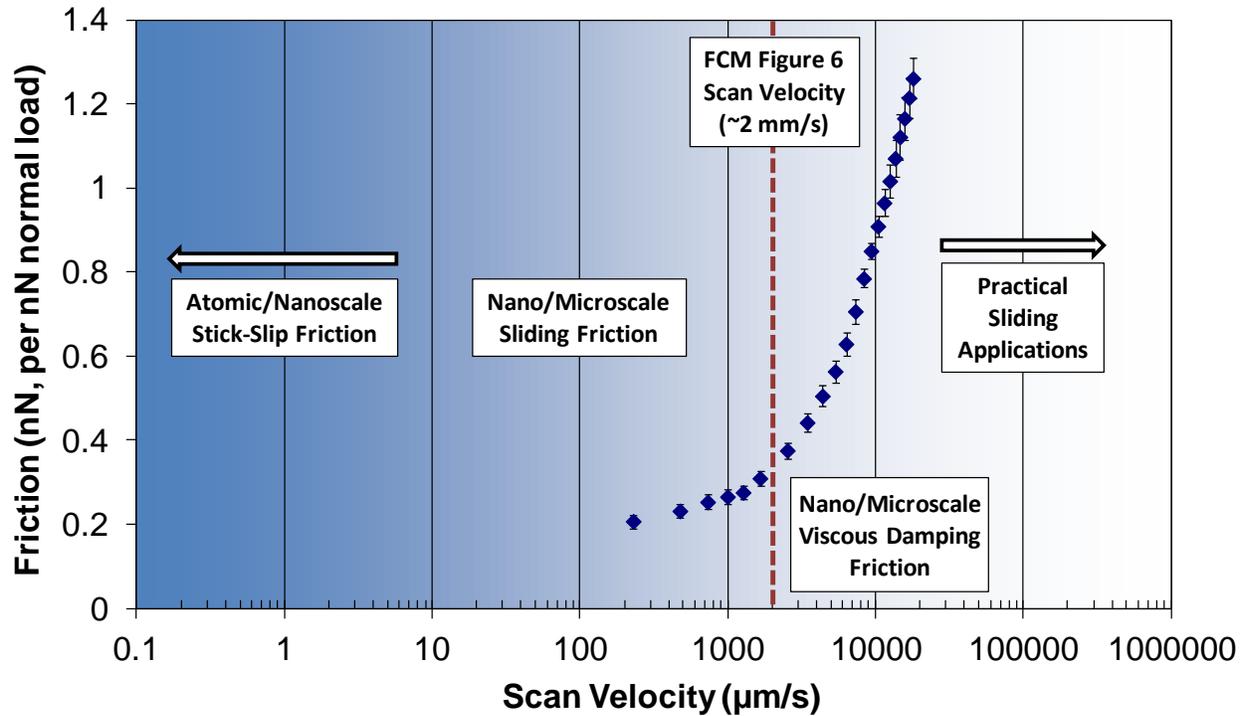

**Figure 8. Friction Force normalized by normal load, versus sliding velocity, for a diamond coated probe on a cleaved mica substrate, noting ranges of relevant friction behavior and the speed used for Figure 6. The Friction Forces were recorded at a line rate of 1000 Hz.**

The dominant friction mechanisms for the data points considered in Figure 8 are sliding friction and viscous damping effects. Atomic scale stick-slip mechanisms, on the other hand, are unlikely to play a role [25, 26] unless much lower velocities (and forces) were employed as noted in the plot. Future work specifically investigating the relevant mechanisms, e.g. considering stress-modified thermal activation, would be insightful, but the present effort is focused on the imaging and speed capabilities of FCM in general.

Clearly, for low speeds the friction increases linearly versus the logarithm of scan velocity. This suggests a sliding friction mechanism as indicated in Figure 8, agreeing with previous observations over similar ranges using standard friction microscopy approaches [22-24, 26]. Under sliding friction conditions, the AFM probe is travelling with a low enough sliding velocity to cause adequate interaction with the substrate, where the tip is fixed in the minimum of the interaction potential (the "stick" state). As the probe travels further, the tip "slips" to the next interaction potential minimum. During this period, the "slip" velocity of the tip is much higher than the sliding velocity. It is during this "slip" motion that most of the energy is dissipated, and is therefore independent of the sliding velocity of the probe for a wide range of sliding velocities [23]. As the velocity increases further, on the other hand, the friction increases faster, with a higher slope versus the log of velocity that continued to the highest speed achieved in our work, 2 cm/sec. Based on modeling or non-imaging friction measurements, such an increase in friction at higher speeds is generally attributed to viscous damping forces [10, 23, 24]. During this regime, the tip has a reduced ability to displace adsorbed molecules on the substrate, resulting in higher friction force. Such effects have been reported to be stronger for scanning in inert atmospheric conditions [27], but the experiments performed here up to substantially higher speeds display the viscous damping effect under ambient conditions as well.

With the continued development of small and/or higher precision lateral or rotational actuators, such friction mapping and variable speed measurements are clearly useful for investigating the friction at practically relevant sliding velocities.

## 5. Friction Mapping Artifacts

Three primary artifacts are important to consider in terms of error in any friction measurements. First, the tip may not properly maintain a constant normal load in all locations. This would cause the locally applied normal force to be different from that anticipated based on the simple AFM setpoint value, shifting the affected friction points laterally on their corresponding friction versus normal force curves (Figure 4). This can easily be accounted for, though, by simultaneously recording the normal deflection along with the torsional (LFM) signal during scanning, in fact as performed here. As a result, the precisely known normal load is incorporated at every image pixel, for every frame in the montage of Figure 5, therefore yielding more precise friction–force curves throughout the imaged area. One practical consequence is that the overall range of normal loads differs somewhat from location to location. Still, it is trivial during analysis to simply consider a uniform loading range for each image pixel when calculating maps of the friction coefficient (slope of each pixel's force-friction curve) or other friction parameters.

This correction does not account for the second category of artifacts, however, those due to changes in topography [28] and contact area [29] during scanning. Indeed, edge effects are clearly present at the pit circumference where topographic discontinuities are greatest (the nearly continuous bright ring around the pit). This has been attributed to the ratchet mechanism of friction, and also due to additional torsion created by tip collision with upward sloping asperities, neither of which can be corrected by the standard "trace minus retrace" friction compensation [28]. Such a variation in friction due to the ratchet mechanism is proportional to the slope of the topography, which reaches a maximum of 5° and 12° for the $SiO_2$ and Au regions, respectively, and a maximum of 37° at the $SiO_2$/Au interface. The maximum corresponding friction variations due to the ratchet mechanism (Equation 3 in [28]) are therefore predicted to be 1%, 4%, and 56% for the $SiO_2$, Au, and $SiO_2$/Au interface, respectively. Much greater variations are detected within in all three regions, however, related to the distinct material responses within the $SiO_2$ and Au regions, and caused by tip collision effects at the topographic interface. Such collision effects are difficult to quantify as noted in [27] since they nonlinearly depend on several factors including applied normal load, scan velocity, and tip geometry.

Additional crescent shaped features, with apparently enhanced friction to the right of the pit and depressed friction at the left, are also visible. Their symmetric but opposite contrast evidences their origin: the inherent necessity in friction imaging of subtracting lateral signals from opposite scanning directions, combined with the fact that the normal and hence lateral loads are slightly different when approaching versus just climbing out of the pit at high speeds. Such effects can be diminished with sharper AFM probes or slower scanning, respectively, but are generally unavoidable in all variations of AFM. For flat surfaces, on the other hand, friction artifacts from topography and contact area are expected to be negligible (e.g. the cleaved mica substrate in Figure 8). Any heterogeneity in the friction contrast should therefore be material dependent under these conditions.

The third common artifact source, tip and/or sample wear, is also a general challenge in SPM, conceivably worsened by high speed imaging as employed here. In the present measurements, the highest normal loads are applied first with subsequent images employing consecutively lower loads. Therefore, any sample damage will predominantly occur in the first, high-load image(s).

Wear of course is a progressive phenomenon, but since it is load dependent it will predominantly occur in the first images as well, likely explaining the nonlinear response observed in the high-load early images of the traditional friction measurement of Figure 4. For the tip, the poly-crystalline diamond coating ensures that minimal blunting will occur after the first high-load images. Corroborating these assumptions, the topography images and cross-sections from the dataset analyzed to construct Figures 5-7 show no appreciable changes, even for the finest features. This implies that the tip and sample are stable under such conditions. Certainly, sub 20 nm features are consistently resolved throughout the multiple images that are compiled to generate Figure 6.

## 6. Conclusion

This work discusses a new Atomic Force Microscopy based method enabling areal Friction Coefficient Mapping (FCM) with nanoscale spatial resolution. As presented, the FCM method leverages high speed SPM imaging at 4 full frames per second. In general, however, FCM is applicable with any speed of AFM imaging, simply requiring more patience and drift stability (or corrections) for standard AFM conditions. For example, an experiment with equivalent force and pixel resolution as that in Figure 6 would require a tolerable but inconvenient 117 minutes (1000 times longer) for 1 Hz scan rates. The Friction Coefficient Mapping presented here is therefore a widely applicable advance, compatible with future as well as legacy AFM instruments.

Results acquired for a mica substrate support a transition between two friction regimes as sliding velocities vary from 200 to as high as 20,000 μm/s. Meanwhile, the ratio of friction coefficients between distinct phases in a nanostructured Au and $SiO_2$ test specimen remained equal over 2 orders of magnitude of tip speeds. FCM therefore allows novel friction studies of heterogeneous surfaces, at velocities ranging from traditional speeds to those approaching realistic sliding or rolling applications (cm/s). The velocity dependence of discrete components can further be investigated to understand the influence of distinct phases, defects, interfaces, and/or topographic features, of growing importance for realistically heterogeneous surfaces in applications such as MEMS/NEMS, bio-materials, and data storage systems.

## 7: Experimental

All experiments are performed at room temperature in ambient air using an Asylum Research Cypher AFM. The AFM's internal feedback system is employed throughout in order to try to maintain a constant normal force between tip and surface via the built-in proportional-integral-derivative (PID) controller feedback loop. For optimal speed, however, all other AFM functions are performed externally based on a National Instruments PXIe-1062Q chassis implementing custom National Instruments LabVIEW code. This includes recording both normal (deflection) and lateral (torsion) cantilever signals with a PXIe-6124 acquisition card (4 channel, simultaneous sampling, up to 4 megasamples/s). A PXI-5421 arbitrary waveform generator (16 bit, up to 100 megasamples/s) is also used to externally drive the X and Y piezoactuators of the AFM, synchronized to the data acquisition board via an 80 MHz clock.

During highest speed scanning, performed 'open loop' (i.e. without position feedback), the actual scanning amplitude (image size) and phase (image registry) will depend on the resonant response of the actuator for the fast scan direction. To accommodate this, the lateral scales of all topography, normal deflection, and lateral friction images are calibrated post-imaging based on simultaneously acquired position sensor data, all with respect to closed loop, slow speed images acquired over known distances on calibration standards.

Diamond coated silicon cantilevers (Nanosensors, CDT-FMR-8) are employed throughout, with a quoted tip length of 10-15 μm, cantilever length of 225 ± 5 μm, and resonant frequency of 60-100 kHz. Each cantilever's spring constant was calibrated in situ, following the "wedge method" common for lateral spring constant calibration [19]. This method incorporates the normal spring constant (determined in situ via the widely employed thermal-tune method) [30], normal and lateral sensitivity of the detecting quadrant photodiode, and ratio of normal to lateral forces when scanning sloped surfaces. The calibration specimen is a MikroMasch TGG01 characterization grating, with precise surface slopes defined by exposed Si {111} planes [31]. Typical measured values of the lateral spring constant are $80.6 - 90.9$ N/m, the normal spring constant is $5.5 - 6.2$ N/m, the normal sensitivity ranges from 250 - 295 nN/V, and the lateral sensitivity is 4280 nN/V. Since the various specimens studied are relatively smooth (mica, microfabricated semiconductor structures), imaging is assumed to be achieved with a single asperity protruding from the nanoscale roughness of the diamond coated tips.

For Friction Coefficient Mapping (Figure 6), the 28-image sequence was acquired at 4 frames per second. Drift of the imaged area is therefore minimal during the 7 second experiment, a substantial benefit for high speed SPM. Nevertheless, since the analysis assumes identical locations for any given pixel in every image, standard drift correction algorithms were employed using common image processing software (ImageJ, Matlab). Tracking topographic features, this yielded a linear drift for the entire experiment of just 32 nm and 44 nm (8 and 11 image pixels) in the fast and slow scan directions, respectively. Pixel shifting to align the consecutive images, and truncating any pixels that therefore were not imaged throughout the experiment, leads to final friction maps with 53,352 points (242 by 239 pixels).

Between each frame, the normal force was decremented by ~31 nN in normal force without stopping the scanning process. The vertical feedback loop transitioned within at most 12 milliseconds (12 scan lines), though friction loops are stable throughout this process. There is no impact on the calculated friction coefficient, however, since it is based on the actually measured normal and lateral forces for every pixel. The normal forces were applied from highest to lowest so that any abrupt damage, or gradual wear, for tip or sample occurs primarily in the first (highest load) imaging frames, and therefore does not introduce appreciable error into the multi-image procedure and analysis.

Due to the large normal load range implemented in the experiment, cantilever torsion and its impact on pixel registry has also been considered. The maximum cantilever torsion based on all factors in this experiment (scan size, tip geometry, normal/lateral force, etc.) is just 1 pixel along each scan direction [32]. This corresponds to a registry error of at most 2 pixels since friction is measured by relating trace (-1 pixel) and retrace (+1 pixel) scans. This misregistry has been accounted for where applicable in the image analysis, but is practically negligible.

The coefficient of friction is calculated for each pixel based on the local linear fit of the friction versus the normal force, using the least squares method. Since the results are spatially resolved, the 95% confidence interval for the coefficient of friction, accounting for data scatter, can also be uniquely visualized, Figure 6(b). There is an average 95% confidence interval of 0.02 ± 0.01 in the $SiO_2$ region of interest, and 0.03 ± 0.01 in the Au region of interest. The coefficient of determination, $R^2$, has also been calculated from each linear fit and is compiled in Figure 6(c). The average $R^2$ values for the Au and $SiO_2$ regions of interest are 0.76 ± 0.15 and 0.74 ± 0.11, with a corresponding mode of 0.87 and 0.81, respectively.

The fact that the coefficient of determination is less than 1 could result from either random scatter for each pixel's friction-force data, or from a poor linear fit due to a nonlinear actual response. Accordingly, the skewness of the data to the linear fit has also been calculated for each pixel. Histograms of this skewness for the Au and $SiO_2$ regions are Gaussian peaks with averages and standard deviations of 0.01 ± 0.51 and -0.02 ± 0.32, respectively. Spatial preferences for the skewness are also negligible excepting edge effects (i.e. any given pixel can skew slightly positive or slightly negative, independent of Au, $SiO_2$, etc.). Since the skewness is evenly but randomly distributed about zero, this confirms random scatter around the linear least squares fits at each pixel. Practically, this means that acquiring more images at distinct normal loads would improve the coefficient of determination ($R^2$) and 95% confidence for each pixel. But this would yield only marginal real benefits since the crucial parameter, the magnitude of the friction coefficients, will remain essentially unchanged.

For velocity dependent friction (Figure 8), only the central 50 pixels from each image line (each speed) are used for calculation to conservatively avoid any edge effects as the tip accelerates or decelerates during highest speed scanning. Since the friction data is acquired along 250 scan lines, 12,500 pixels are thus analyzed for each point in the plot.

The patterned SiO$_2$/Au specimen was prepared by Sparse Colloidal Lithography using monolayers of polystyrene (PS) colloids prepared by electrostatic self-assembly as a shadow-mask, described thoroughly elsewhere [33, 34]. Briefly, a Si wafer substrate was sputter coated with 30 nanometers of Au (2nm Ti adhesion layer). The surface was functionalized with a polyelectrolyte triple layer by sequential deposition of positive PDDA (poly(diallyldimethylammonium chloride), negative PSS (poly(sodium-4-styrenesulfonate)) and positive PAX (polyaluminium chloride) monolayers by electrostatic self-assembly giving a stable positive charge at neutral pH. An array of 300 nanometer colloidal PSS spheres was then assembled from solution forming a short range ordered array of particles with spacing determined by the electrostatic repulsion between already adsorbed and later arriving colloids. After a thermal treatment step in water to prevent rearrangement of the film by capillary forces during drying, a polycrystalline SiO$_2$ layer is deposited by electron beam stimulated thermal evaporation. The PSS colloids are subsequently removed by tape striping, leaving a short range ordered array of ~300 nm pits in the SiO$_2$ layer with Au at their base.

**Acknowledgements**

Support by Jeff Honeyman, Anil Gannepalli, and Jason Bemis of Asylum Research is greatly appreciated for custom PID development. BDH appreciates support from the Velux Foundation and iNANO, JB from NSF:DMR:Ceramics, 0909091, and SL from NSF:DMR:IMR, 0817263.

# Figure Captions:

**Table of Contents:**

A new approach is presented for true quantitative mapping of friction with AFM. Although applicable with any legacy system, 'FCM' is demonstrated at high speeds (mm/sec) with results comparable to standard (µm/sec) data. Friction at sliding velocities as high as 2 cm/sec are also reported, evidencing the potential for nanotribology studies at speeds relevant to MEMS, biological implants, etc.